%% file: master.tex
\documentclass[preprint,journal]{vgtc}       

\ifpdf
  \pdfoutput=1\relax                   
  \usepackage{graphicx}                
  \DeclareGraphicsExtensions{.pdf,.png,.jpg,.jpeg} 
\else
  \usepackage{graphicx}                
  \DeclareGraphicsExtensions{.eps}     
\fi%

\graphicspath{{figures/}{pictures/}{images/}{./}} 

\usepackage{todonotes}
\usepackage{microtype}                 
\PassOptionsToPackage{warn}{textcomp}  
\usepackage{textcomp}                  
\usepackage{mathptmx}                  
\usepackage{times}                     
\usepackage{cite}                      
\usepackage{tabu}                      
\usepackage{booktabs}                  
\usepackage{subfigure}
\usepackage{amsmath}
\usepackage{url}

\usepackage{enumitem} 

\onlineid{0}

\vgtccategory{Research}
\vgtcpapertype{Evaluation}

\title{The Effectiveness of Interactive Visualization Techniques for Time Navigation of Dynamic Graphs on Large Displays}

\author{Alexandra Lee, Daniel Archambault, and Miguel A. Nacenta}
\authorfooter{
\item
 Alexandra Lee is with the Dept. of Computer Science, Swansea University and Swansea University Medical School, UK.\\E-mail: a.s.lee@swansea.ac.uk
\item
 Daniel Archambault is with the Dept. of Computer Science, Swansea University, UK.  E-mail: d.w.archambault@swansea.ac.uk
\item
 Miguel A. Nacenta is with the Dept. of Computer Science, University of Victoria, Canada. E-mail: nacenta@uvic.ca
}

\shortauthortitle{Lee \MakeLowercase{\textit{et al.}}: Interactive Visualization Techniques for Dynamic Graphs on Large Displays}

\abstract{
Dynamic networks can be challenging to analyze visually, especially if they span a large time range during which new nodes and edges can appear and disappear. Although it is straightforward to provide interfaces for visualization that represent multiple states of the network (i.e., multiple timeslices) either simultaneously (e.g., through small multiples) or interactively (e.g., through interactive animation), these interfaces might not support tasks in which disjoint timeslices need to be compared. Since these tasks are key for understanding the dynamic aspects of the network, understanding which interactive visualizations best support these tasks is important. We present the results of a series of laboratory experiments comparing two traditional approaches (small multiples and interactive animation), with a more recent approach based on interactive timeslicing. The tasks were performed on a large display through a touch interface. Participants completed 24 trials of three tasks with all techniques. The results show that interactive timeslicing brings benefit when comparing distant points in time, but less benefits when analyzing contiguous intervals of time. 
} 


\CCScatlist{ 
 \CCScat{K.6.1}{Management of Computing and Information Systems}%
{Project and People Management}{Life Cycle};
 \CCScat{K.7.m}{The Computing Profession}{Miscellaneous}{Ethics}
}

\vgtcinsertpkg

\begin{document}


\firstsection{Introduction}

\maketitle
\newcommand{\expnumber}[2]{{#1}\mathrm{e}{#2}}

\input{sections/01-Intro}
\input{sections/02-RWork}

\input{sections/03-Interfaces}
\input{sections/04-ExperimentalApproach}

\input{sections/05-Experiment1}
\input{sections/06-Experiment2}
\input{sections/07-Experiment3}
\input{sections/08-Discussion}
\input{sections/09-Conclusion}

\acknowledgments{
We would like to thank the EPSRC grant EP/N509553/1. This work was partially supported by a Microsoft Surface Hub award.}

\bibliographystyle{abbrv-doi}

\bibliography{master.bib}
\end{document}

%% file: sections/01-Intro.tex
Dynamic networks are networks that change over time. Nodes and links might appear or disappear at different points in time and attribute values may change. Dynamic networks appear in many domains including social science~\cite{brown_fischer_goldwich_keller_young_plener_2017}, transportation~\cite{gallotti2015multilayer}, digital communications~\cite{gloor2004tecflow}, epidemiology~\cite{masuda2013predicting}, and others. 
These networks are difficult to analyze and interpret and can therefore benefit from having interactive visualization techniques applied to them.

Dynamic networks are most commonly visualized by two approaches~\cite{muller_visualization_2003,Beck2017,kerracher_design_2014, Bach2017}. One approach is an interactive animated representation where the user can control which moment in time is being displayed. 
The other is to split the time domain into a series of timeslices and represent them separately as small multiples. This latter approach is currently the most popular in the literature. Multiple studies have shown that the small multiples approach is faster than interactive animation with no significant differences in error rate~\cite{5473226,farrugia_effective_2011,Archambault2016}.

The above approaches and experiments all assume uniform slicing at a given level of granularity.  However, what uniform duration of timeslice should be chosen?  If the timeslices are too coarse, the representation collapses too many events onto the same timeslice, hiding the subtlety and the true order of events within each timeslice. If the timeslices are too fine, there are too many points in time to navigate in the data and analysts will have a hard time remembering timeslices that are off screen when identifying patterns. Interactive timeslicing addresses this issue by allowing the analyst to interactively select the width and location of timeslices with the possibility of representing several of these timeslices at once~\cite{19LeePlaid}. Similar methods have been used in other areas~\cite{06plumlee}, but the effectiveness for interactive timeslicing for dynamic graphs is still unknown.

We designed a series of experiments to test an interactive timeslicing approach against interactive animation and small multiples for navigating time in dynamic graphs. 
We tested these approaches on a touch-based 84" display with 4K resolution that we consider representative of advanced visualization set-ups in current collaborative professional settings~\cite{19LeePlaid}. We discuss generalizability implications in Section~\ref{sec:limitations}.
Currently, there is no evidence that an interactive timeslicing approach will be better; the additional complexity of interaction might be too costly, in terms of time, which could negate all benefits.

We conducted three experiments \footnote{All experimental material is available on the OSF at \url{https://osf.io/bdpnr/?view_only=8d2e29693b714b7d8bb4abd407ad8e56}.} where participants interactively navigated time to find: a) changes in graph structure at points in time (specific timeslices), b) changes in graph structure across an interval of time (a series of consecutive timeslices), and c) changes in attributes at points in time.  For each experiment we had two conditions:  near, where the moments in time of interest were close to each other in time, and far, where the moments were further apart in time. 

 Our results show that interactive timeslicing shortens completion time (small multiples takes $1.42$ times longer) and improves correctness ($5$ percentage point improvement over small multiples) for comparing graph structure at discrete time points. The differences increase almost by a factor of two for temporally distant network events.  For finding attribute changes within discrete time points, interactive timeslicing improves completion times (small multiples takes $1.19$ times longer for near time intervals). 
However, for finding graph structure changes over time intervals, small multiples outperforms interactive timeslicing (interactive timeslicing takes $1.42$ times longer). 
 These results show how the addition of a relatively simple interactive feature can greatly facilitate the challenging and important analysis tasks of dynamic networks. Our results also further generalize previous findings from the literature on additional tasks that found small multiples has faster completion times when compared to interactive animations for graph analysis with no difference in correctness.       

%% file: sections/02-RWork.tex
\section{Related Work}
We review related work for dynamic network visualization techniques along with relevant empirical evaluations. We also provide an overview of network visualization techniques on large displays.

\subsection{Dynamic Network Visualization Techniques}
Beck et al. \cite{Beck2017} separate dynamic graph visualizations into two categories based on their method for encoding time. Time-to-time mappings represent time naturally via the temporal dimension, the most common example of this being animation. Time-to-space mappings use one or more spatial dimensions to encode the temporal information, a common example of this is small multiples.  

\paragraph*{Time-to-Time Mapping}
Animation was one of the first, and is still one of the most common, approaches for visualizing dynamic data.  In interactive animation, an interactive video plays a movie of the evolving graph. The result is reasonably intuitive for node-link visualizations of data (e.g., ~\cite{6658746, 1382908,08Frishman}). 
However it requires the user to heavily rely on his or her memory as multiple time points are not visible concurrently.  Thus, frequent backwards and forwards navigation through the data is required, incurring interaction costs~\cite{5473226} and additional errors due to fatigue and reliance on memory.

\paragraph*{Time-to-Space Mapping}
Timeline-based visualizations encode the temporal dimension as one or more spatial dimensions. These visualizations can be broken down into four categories: node-link based approaches \cite{6065001,Shneiderman_task_2006}, matrix based approaches \cite{brandes_asymmetric_2011,yi_timematrix_2010, bezerianos_graphdice:_2010, y2008improving}, hybrid approaches \cite{hadlak_situ_2011}, and comic-style approaches~\cite{bach2016telling}. Additionally, scalable line charts can effectively show the variations in an attribute value over time~\cite{walker2016timenotes}.  Timeline-based visualizations have the advantage of displaying multiple timeslices on screen simultaneously.  However, when event order within a timeslice is essential it is hard to draw conclusions about these events~\cite{8666650}. Also, if the timeslices are separated by large spans of time, interaction is required to scroll back and forth between them for comparison when reordering the timeslices is not possible.

\subsection{Experimental Evaluations and Dynamic Data}
Much existing work has evaluated the performance of different time and visualization types within a mapping. 

Saraiya et al.~\cite{1532151} compared time-to-time and time-to-space mappings for node-link diagrams when interacting with multidimensional data. Animation performed better for two points in time, but tasks involving more time steps were better served by timeline-based approaches.  A number of studies have demonstrated advantages for time-to-space mappings on dynamic graphs~\cite{5473226,Archambault2016, farrugia_effective_2011} and visualization in general \cite{tversky2002animation}.  
These studies use linear interpolation (fading nodes in/out from the visualization) and find that small multiples is faster with no significant differences in terms of error rate.  Boyandin et al.~\cite{boy} find that animation facilitates findings on adjacent time steps but that small multiples allow the discovery of patterns which last for greater periods of time. Hybrid approaches mixing animation and timelines can, under certain conditions, produce better results than animation or timeline approaches alone~\cite{rufiange_diffani:_2013}.  

The majority of these experiments have focused on the structural properties of the network first and the time navigation second.  Also, all the above experiments assumed a uniform timeslicing selected beforehand whereas many recent techniques for graph visualization do not make this assumption~\cite{19LeePlaid,Dynnoslice,DynnosliceTVCG,19Wang}.  For dynamic graphs that are long in time (for example, events lasting seconds over months of data), no experiments have been run.  Also, interactive timeslicing has not been evaluated.  We present three experiments that evaluate user interaction with the time dimension for long in time dynamic graphs on a large touch display.

\subsection {Visualization on Large-Screen Devices}
Visualization on large displays has a long history and is appealing for a variety of reasons. In the past, larger displays (usually composites of many smaller displays~\cite{streitz_roomware_1998,guimbretiere_fluid_2001,ball_analysis_2005,tao_ni_survey_2006,bi_comparing_2009}) were, due to technology constraints, the easiest way to increase the available pixel count. This, in turn, increased the ability to display detail or more data items. Increased number of pixels and larger size has been shown to increase performance and has perceived benefits~\cite{czerwinski_toward_2003,tan_physically_2006,tan_physically_2004}.

Although modern display technology has reached pixel densities that make the argument about pixel counts largely irrelevant (a state-of-the-art 15'' display can have as many pixels as an 88'' display from just a few years ago), there are still reasons why large displays are desirable for visualization. First, they enable larger numbers of people to work on the same data~\cite{jakobsen_up_2014}. Second, people interacting with large displays seem to benefit from physical navigation~\cite{ball_move_2007,jakobsen_is_2015, andrews_impact_2013}, which might improve memory and performance~\cite{jakobsen_up_2014}. The ability to change the distance to the display easily and naturally by stepping back and forth also enables a natural zoom experience and supports visualization techniques that would be hard or awkward with personal displays (e.g.,~\cite{jakobsen_is_2015,isenberg_hybrid-image_2013,nacenta_fatfonts_2012}). Incidentally, a larger display will also reduce problems with the precision of touch interaction (e.g., the fat finger problem~\cite{siek_fat_2005}) simply because the content will normally be larger. 

Perhaps due to these reasons, research investigating and translating visualization to large displays is still active (e.g.,~\cite{7516722,kister2017grasp,19LeePlaid}). From the information available in literature and our own experience, we speculate that large displays are likely to offer the best environment for the visualization of complex and high-density information such as dynamic networks, particularly when scalability in the time dimension is required.  The specialized nature of this kind of analysis also means that the extra cost of procuring large displays is usually well justified, and many research environments already offer medium to very large displays. Hence, we carried out our implementation on a relatively large display that matches the use of dynamic network visualizations that we envision in the near future. Our implementations do not include any interactive or visual features that prevent use with a small personal display. Therefore, we do not expect that the comparisons between human performance or preference between the techniques would vary in a smaller display, although this will need to be supported empirically in the future.

%% file: sections/03-Interfaces.tex
\section{Experiment Interfaces}
\label{label:interfaces}
For this experiment, we consider three dynamic graph visualization techniques: small multiples, interactive animation, and interactive timeslicing. Animation and small multiples were chosen because they are widely considered the dominant alternatives in dynamic graph visualization \cite{Beck2017}, are in common use, and have been revisited in recent data visualization experiments for small displays~\cite{Brehmer2020}. Interactive timeslicing represents a novel alternative that is promising but has not yet been compared with the other two approaches~\cite{19LeePlaid}. 

Node-link diagrams were used for all graph representations as they are popular in media and are well explored in literature. All approaches were implemented to work on a large touch-enabled display (see Apparatus).


\subsection{Interactive Animation}
Interactive animations present the dynamic graph as an interactive film, with the user being given control of playing the dynamic data via a slider that can travel both forwards and backwards in time. Nodes and edges fade in and out of the drawing area as they are inserted or removed from the network.

\begin{figure*}
    \centering    
    \includegraphics[width=0.9\textwidth]{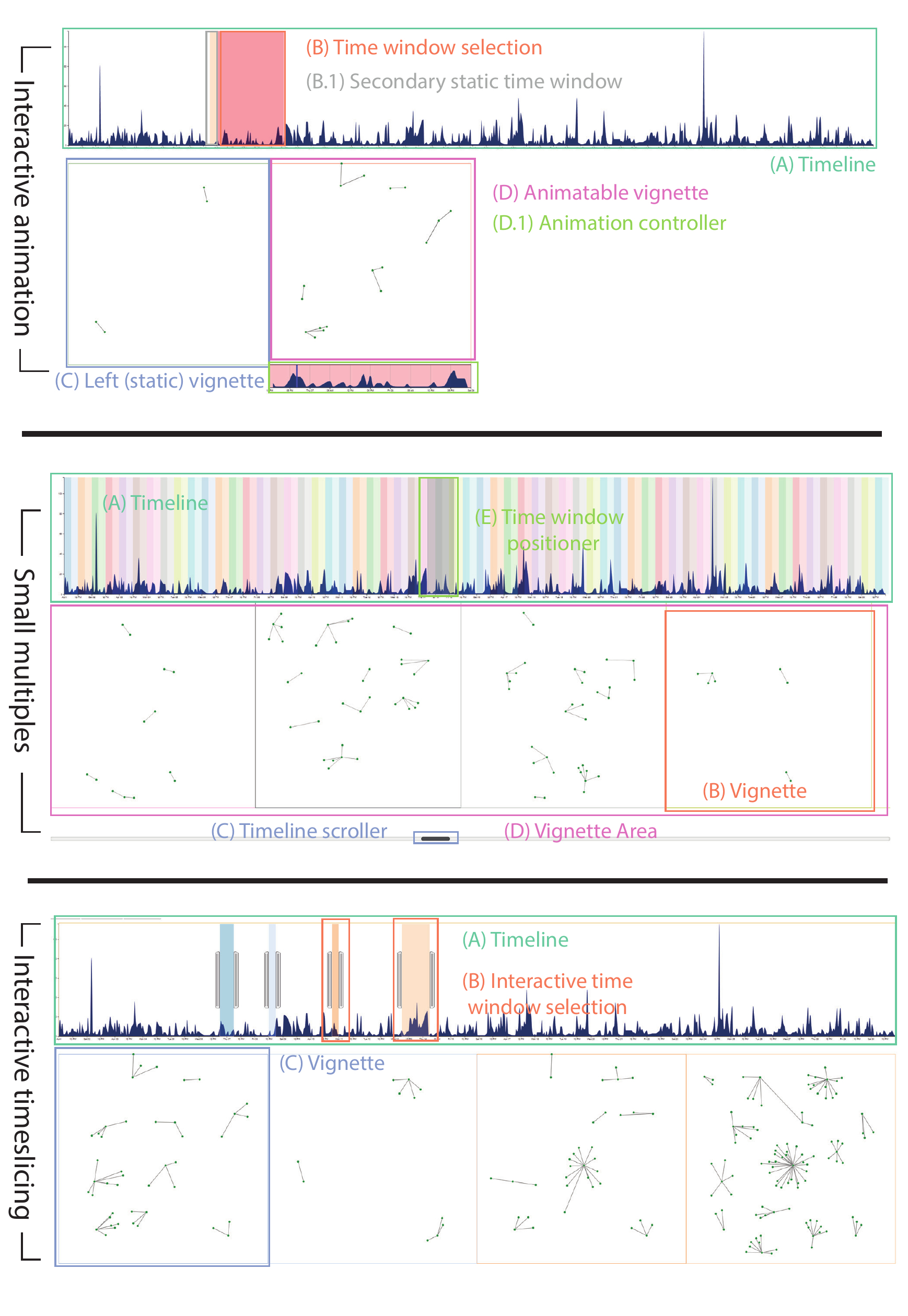}
    \caption{The interfaces used in this experiment.  Interactive animation is at the top, small multiples in the middle, and interactive timeslicing at the bottom.}
    \label{fig:interface-all}
\end{figure*}

For this study our animation interface has four components. The component labeled (A) (see Figure~\ref{fig:interface-all}, interactive animation) was the main timeline which showed the number of edges within the dataset aggregated at the hour level. The component labeled (B) was an interactive time window selection. Users would touch and drag to select a time window from the longer time series, with component (B.1) also selecting the 6 hour time period immediately preceding the user selected time window. An alternative  solution would have been to animate directly on the long timeline.  However, this could potentially introduce a confound into our experiment as the animation would need to consider a much larger amount of data (the entire long-in-time dynamic graph) rather than a shorter animation around a given time window.  Therefore, we decided to allow the user to select the animation window on the longer timeline around the target area.

Component (C) is the flattened graph representation of the time window selected by component (B.1). This static graph allows the participant to enter an answer without navigating through the network. Component (D) is the interactive animation window --- when the time range was first selected, component (D) showed a flattened, static, representation of the graph within the selected time range. The participant pressed and dragged on area (D.1) to control the rate and position of the animation. A purple line within (D.1) follows the participant's finger to show the current temporal graph position relative to the timeline.

\subsection{Small Multiples}

Small multiples uses many interrelated graphs to represent time. It is analogous to a comic book representation for time where the evolution of the data can be tracked by reading across the frames from left to right.  We use the term small multiples as it is most similar to those in other experiments~\cite {Archambault2016,5473226,farrugia_effective_2011,Robertson2008,Brehmer2020}.  For temporal networks, it is common to uniformly timeslice data by flattening each timeslice into its own window, with these graphs then drawn next to each other in a timeline pattern. Small multiples can be used for most forms of dynamic data; in dynamic graph visualization, it can be used both with matrices and node-link diagrams.

For the small multiples representation in these experiments, our study dataset (Section~\ref{sec:dataset}) was divided into blocks of 6 hours, giving a total of 120 timeslices which were translated into vignettes. Along the top of the screen, labeled (A), is the timeline. Each block of color defines one 6 hour time range, and each block of color also corresponds to the border color of the relevant vignette shown in (B).
The vignette area was designed to only show four small multiple representations simultaneously due to screen size and readability constraints. The remaining vignettes could be accessed by moving the scroll bar at the bottom of the screen, (C), touching and dragging on the vignette area, (D), or by moving the light gray time-positioner on the timeline, (E).
The gray time-positioner, (E), helped participants map the current position of the vignette area to its temporal position within the entire time series without performing color comparisons of the vignette borders.

\subsection{Interactive Timeslicing}

This interface was based on a previous interactive timeslicing design~\cite{19LeePlaid} where users could define custom time ranges of any desired size or order and have the individual representations positioned based upon the order of user interactions. 

Component (A) is the timeline that is common to all of our experimental interfaces. Participants touched and dragged at any position on the timeline to create a new time window selection of any size (B).  A time window selection created a new vignette (C) containing a flattened, static, representation of the graph. A time window selection could be any multiple of 6 hours, with the edges of the selection snapping to the same timeslice intervals as all conditions --- this was to avoid small off-by-one errors that may seriously hinder the ability of a participant to successfully complete any given task.

The maximum number of interactive timeslices that could appear onscreen at any time was four. This limitation was introduced to avoid giving interactive timeslicing an artificial advantage over small multiples. Interactive selection of timeslices in this way allows the participant to make independent time windows at very distant points in time for the network.  However, there is an interaction cost associated with creating the time windows that is higher than both animation and small multiples:  the location in time needs to be determined first and then the participant needs to draw a much more precise window both in terms of its width and position.

%% file: sections/04-ExperimentalApproach.tex
\section{Experiment and Research Questions}\label{sec:questions}
Our general goal is to provide empirical evidence that can support the design of better interfaces for time navigation when exploring dynamic networks. 
Thus, we designed three experiments, each testing a different task. For each experiment we seek to answer the following questions:

\begin{enumerate}[label=Q\arabic*]
    \itemsep0em 
    \item Which interface will have the lowest completion times for the selected tasks?
    \item Which interface will be the most accurate for completing the selected tasks?
    \item How does distance in time between data elements of interest affect the performance with the different interfaces?
\end{enumerate}

\section{Experimental Approach}
We designed a series of three experiments through the following iterative process. First, we looked at existing relevant taxonomies of tasks~\cite{Andrienko_task_2005,kerracher_task_2015,14Ahn} and at findings from previous observations of analysts working with dynamic networks~\cite{19LeePlaid}. Then we filtered out tasks that do not involve navigation between at least two given time points. We excluded tasks that only involve single time points, since are we primarily interested in challenges specific to dynamic, rather than static, networks. From the remaining tasks we selected three tasks with the following criteria: a) we preferred low-level tasks that might be components of larger tasks, and b) we preferred tasks that were very different from each other and where the focus was on different elements of the data (e.g., the structure of the network or the variation of attributes) and how they varied over time.

The chosen tasks can be broadly summarized as:
\begin{enumerate}[label=E\arabic*]
    \itemsep0em
    \item Detecting graph structure changes at discrete points in time.
    \item Detecting graph structure changes over a time interval.
    \item Detecting single attribute change at discrete points in time.
\end{enumerate}

We run the three experiments with the same participants in the same session.We prioritize comparing results across interfaces on the same task; to reduce noise from order effects on our comparisons of interest, we run the three experiments in the same order for all participants (i.e., we do not randomize experiment order). Experiments took place in early March 2020 and took approximately 85 minutes to complete.

\subsection{Participants}
Twenty-four unpaid volunteers (8 identified as females, and 16 as males, age range of 18 to 32) from our department participated in the study.  The number of participants was decided in advance using our prior experience from the four pilots and to ensure proper counterbalancing.
We used a simple questionnaire to screen participants without typical color vision and those who did not have basic computer and mathematics experience. We did not require prior knowledge of networks. 

Before beginning the first experiment, participants filled a short demographic questionnaire indicating their familiarity with networks and network graph visualization using a 1--5 Likert scale. 
Four participants gave their familiarity as 1 (no knowledge), six as 2, nine as 3, three as 4, and two as 5 (high knowledge).

\subsection{Apparatus}
The interfaces were built in JavaScript, primarily leveraging the D3.js library \cite{D3} with some modifications. Conditions ran in a Chromium browser window which was wrapped in a QT front end interface, with a Python Flask back-end. Implementations of the interfaces are identical across the three experiments; visual indicators for the task and methods to indicate the answer vary with the task and are described within each experiment. 

All three of these approaches were implemented for an 86" wall display with 4k resolution (3840 by 2160 pixels) mounted at 90cm from the floor.

\subsection{Experimental Dataset and Graph Layout}
\label{sec:dataset}
The dataset used in these experiments is a filtered version of a previously collected dataset~\cite{brown_fischer_goldwich_keller_young_plener_2017} which describes interactions between anonymized actors on Instagram who liked or commented on non-suicidal self-injury posts. 
Edges have unique identifiers, have a specific occurrence time (down to second precision), and do not have a duration (i.e., are considered atomic, only measured at the time of posting).  A duration of one hour is assigned to each edge at its posting time to ensure it is visible and lies completely within the six hour time window. Additional attributes of the dataset were removed for anonymity; hence the dataset contains only nodes, edges, and the time that the nodes and edges appeared in the dataset. 

The dataset is representative of typical dynamic networks but was substantially filtered to reduce its size to make our tasks feasible. We removed single interactions between actors (cases where two actors only interacted with each other once), self-interactions, actors that received less than $250$ interactions, and actors who only interacted with removed actors. This allowed a stable graph layout and made task duration appropriate to the available time. The final dataset consists of $776$ distinct nodes and $8182$ edges over a 30-day period of time. It had an approximate average density of 11 events per hour, or 66 events per 6 hour timeslice. As this dataset is based on real data, the event distribution was non-uniform and therefore some time periods had a higher event density, while others had a lower event density.  This variability over long time periods allowed for very different graphs to be presented at the time points and intervals in our tasks, which increases the generalizability of trials.

Our dataset is long in time with a fine grained temporal resolution (hour resolution over a month of data), meaning that there are hundreds of timeslices.  Typical timeslice-based approaches~\cite{11Mader}, in particular linking strategies, generally do not scale to hundreds of timeslices as nodes move unnecessarily with the many inter-timeslice edges being artificially inserted~\cite{Dynnoslice}.  As we wanted to control for graph layout across conditions we could not draw the graph on demand as the user navigated through time. Thus, we needed to draw the full dynamic graph beforehand.
Event-based dynamic network drawing techniques could provide a solution~\cite{Dynnoslice,DynnosliceTVCG} but they only scale to about 5000 events.  Therefore, we computed a fixed layout for nodes and edges that uses pinning strategies so that nodes retain position whenever they appear, allowing the drawing stability to consistently support the mental map across conditions for all experiments.  The final graph layout was generated first with the Visone aggregation strategy~\cite{visone} (all events collapsed down to a single timeslice) and then cleaned slightly in Gephi~\cite{gephi}, using its implementation of Fruchterman and Reingold's algorithm~\cite{fruchterman1991graph}. Although more scalable algorithms are available~\cite{05Hachul}, our graphs are relatively small in terms of the number of nodes and edges. This allows us to use standard force-directed approaches instead of multilevel ones.  The resulting layout still contained some overlapping nodes and edges; we manually adjusted overlapping items.

\subsection{Experimental Design and Procedure}

All three experiments share an identical structure in terms of factors, conditions, and repetitions. The main manipulation of interest is \emph{interface}, with the three interfaces {\tt Anim}, {\tt Mult}, and {\tt Int TS} as levels---see Section~\ref{label:interfaces}. One factor is temporal distance (near and far), which manipulates the separation of the target timeslices or the length of the time interval involved in the task. In the near condition, timeslices had a separation of 18 hours (three timeslices) while in the far condition timeslices had a separation of four days (16 timeslices).

We performed a within-participant design with participants completing 3 repetitions for each cell, for a total of 18 trials, as well as two additional easy tasks on each interface for training which was discarded before data analysis. Participants carried out trials (including training) in three blocks of 8, with 15 second breaks between each block. 
Possible ordering effects were counterbalanced using Latin squares for each participant. A participant always did the interfaces in the same order for each experiment that they completed. Trials in the near condition always took place before trials in the far condition. We did not counterbalance temporal distance because we are not interested in the quantitative comparison of performance between near and far conditions.

Prior to the real trials of each interface condition participants received a tutorial on the specific interface and carried out one trial example.
Upon completion of each experiment participants ranked each condition on a scale of 1 (best) to 3 (worst), according to their preferred interface for completing the type of tasks tested. The experimenter also collected qualitative notes during the experiment. Participants had the opportunity to make other comments and share their thoughts about the tasks, interfaces, and hardware.

\subsection{Statistical Methodology}
The statistical methodology was decided during the experimental design process and recorded before data collection.  All three experiments employed the same methodology.

Correctness and completion time were measured and analyzed separately for all three experiments.  Completion time was measured as the number of seconds (s) to complete each task for all three experiments. Measurements for correctness varied for each experiment, and these measures are detailed in their respective sections. However, for all three experiments, correctness was measured on a $[0,1]$ interval (with $1$ corresponding to $100\%$ correctness).

As we were interested in determining the performance of the three interfaces under near and far conditions, we chose to divide the data of each experiment by the near and far factor before beginning the analysis.  The completion times and correctness of the three repetitions in each cell was averaged per participant.  Completion time was log transformed (log2) before analysis and compared through pairwise, two tailed t-tests.  We were uncertain if the distribution of correctness data would follow a normal distribution for our measurements, as a result we applied a Shapiro-Wilk test with $\alpha=0.05$ to each condition of near and far for each experiment separately.  For all three experiments, the correctness data did not usually follow a normal distribution.  Therefore, two tailed, pairwise Wilcoxon signed rank tests were used for correctness in all three experiments.  For each experiment, six pairwise comparisons (three for near and three for far) for time and six pairwise comparisons for correctness were performed.  Holm–Bonferroni corrections were used to determine significant results.

The results are presented in the next section.  In all result figures, blue corresponds to animation ({\tt Anim}), red to small multiples ({\tt Mult}), and orange for interactive timeslicing ({\tt Int TS}).  The mean is indicated using a circle and the median using a square.  Error bars represent bootstrapped 95\% confidence intervals computed using 250,000 repetitions.  We report p-values to three decimal places in all figures~\cite{19dragicevic,16Greenland,17Greenland,16Lazar}. Solid lines indicate $p < 0.05$ and dashed lines $p \geq 0.05$.

%% file: sections/05-Experiment1.tex
\section{Exp. 1: Graph Structure Changes at Points in Time}
This experiment tests completion time and accuracy for tasks where the structure of the graph had to be compared at two separate time points.  We suspected that interactive timeslicing would perform the best for this task ({\bf Q1} and {\bf Q2}).  In addition, we anticipated that interactive timeslicing would perform much better than animation and small multiples for the far condition. However, we were less sure of the performance of the remaining two interfaces on this condition of the experiment as they have not been tested and were not designed for exploring distant points in time ({\bf Q3}). 


\paragraph {\bf Task and Procedure.}
The task prompt provided to participants for experiment one was `each pair of red (start) and blue (stop) lines signify a timeslice. In the first timeslice, click on all edges that disappear at least once in all other timeslices.' The target timeslices of six hours each began with a red line and finished with a blue line.  Participants completed the task by selecting edges in the first window that did not appear in the other window. 




The video figure in the supplementary material illustrates how participants answered this task on all three interfaces.  For all interfaces, answers to the question were entered through lasso selection on the leftmost vignette to control for answer entry. Edges selected using the lasso tool were coloured red and increased 4$\times$ in width and could be de-selected.

\paragraph{\bf Animation.}
Participants dragged on the timeline starting from the end of the first red-blue pair of timeslice demarcation lines, ensuring that the vignette to define an answer was correctly created.  This action would generate two vignettes: a left vignette displaying the selected time interval of the first pair of timeslice lines as a static, flattened, graph for answer entry; and a right vignette which initially displayed the flattened, static, graph covering the whole time range selected by the participant during the original time selection operation. 

A timeline appeared at the bottom of the right hand vignette displaying a zoomed version of the selected area of the main timeline, with task target timeslice areas shaded in blue. Participants were able to touch and drag on the timeline to animate the right hand vignette. Touching and dragging in the left vignette would activate the lasso tool to allow participants to select, or deselect, edges for their task answer.

\paragraph{\bf Small Multiples.}
Participants used either the grey time positioner (Figure~\ref{fig:interface-all}, small multiples, item E) or the scroll bar attached to the vignette area (Figure~\ref{fig:interface-all}, small multiples, item C) to navigate to the correct position in time. Task relevant vignettes were distinguished from non-task vignettes by increasing the border thickness 4$\times$. Participants were able to touch and drag in the left vignette to use the lasso tool to define the answer set.

\paragraph{\bf Interactive Timeslicing.}
Participants touched and dragged on the correct area of the timeline to create a time window selection (Figure~\ref{fig:interface-all}, interactive timeslicing, item B). To avoid small selection errors these user-created time windows always readjusted to the nearest 6 hour boundary, ensuring that the selection was the exact time period required by the task. 
After the creation of a time window a corresponding vignette appeared below the timeline (Figure~\ref{fig:interface-all}, interactive timeslicing, item C).  Touching and dragging in the left-most vignette with the lasso tool defined an answer to the given task.

\paragraph{\bf Measurements.} Time (number of seconds) and accuracy were measured for this experiment.  Accuracy involved comparing  the set of edges selected by the user and the set of edges present in the correct answer of the question.  In order to evaluate correctness in this case, we employ a method from pattern recognition and information retrieval.  A perfect answer would have perfect \textit {precision} ($p$) (no edges outside the correct answer are selected) and perfect \textit {recall} ($r$) (all of the edges in the correct answer are selected).  Precision and recall can be combined together into the $F_1$ score to give a measure between $[0,1]$:

\begin{equation}
    F_1 = 2\frac{pr}{p + r}
    \label{eqf1}
\end{equation}

The value of $p$ and $r$ are defined in the following way.  Consider two edge sets: the set of participant answer edges ($X$), and the set of correct answer edges ($Y$).
\begin{equation}
        p = \frac{|X \cap Y|}{|X|}
\end{equation}
\begin{equation}
        r = \frac{|X \cap Y|}{|Y|}
\end{equation}

\begin{figure}[t]
    \centering
    \includegraphics[width=0.48\textwidth]{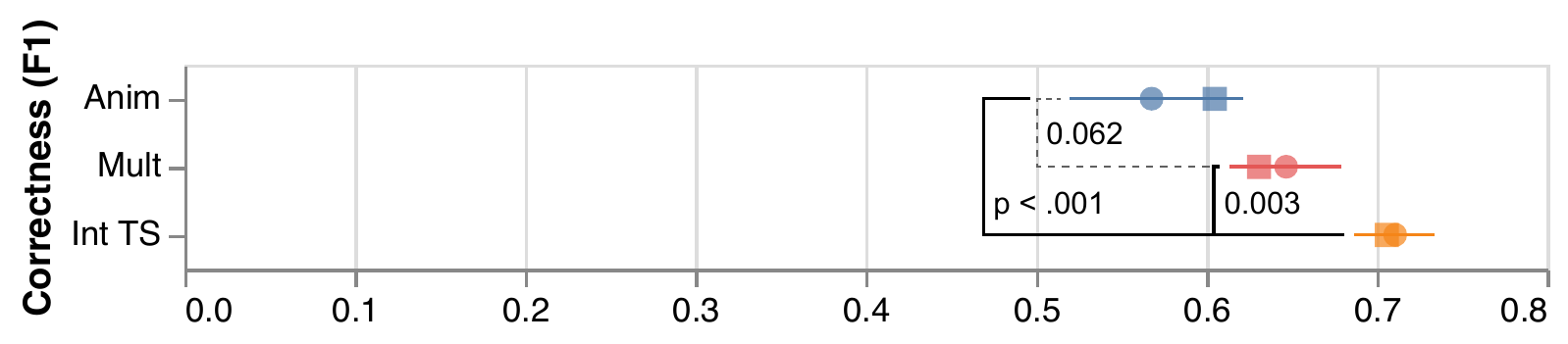}
    \includegraphics[width=0.48\textwidth]{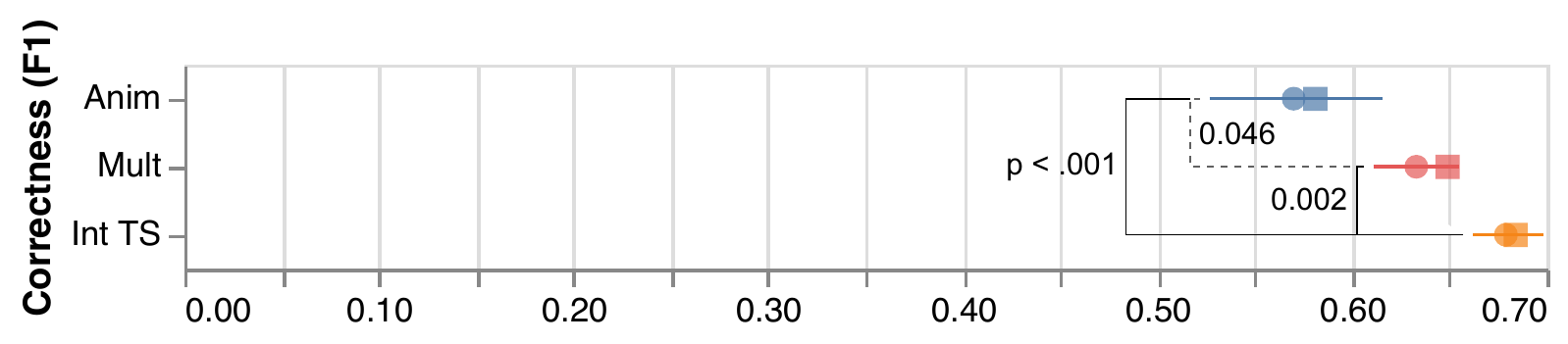}
    \caption{Experiment 1 correctness for the near (top) and far (bottom) conditions as computed by the $F_1$ score (Equation~\eqref{eqf1}). The dot represents the mean, square the median, and 95\% CIs shown.  Pairwise lines indicate p-values with solid lines indicating significant differences.}
    \label{fig:E1Corr}
\end{figure}

\begin{figure}
    \centering
    \includegraphics[width=0.48\textwidth]{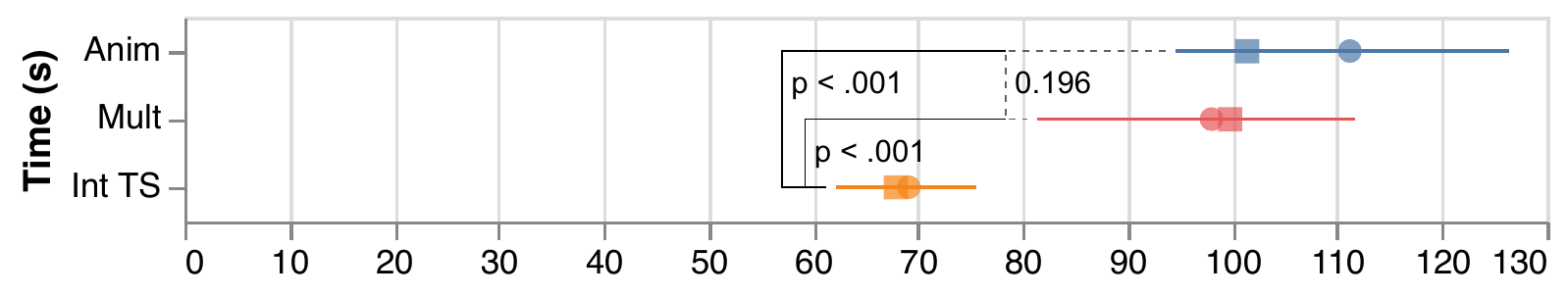}
    \includegraphics[width=0.48\textwidth]{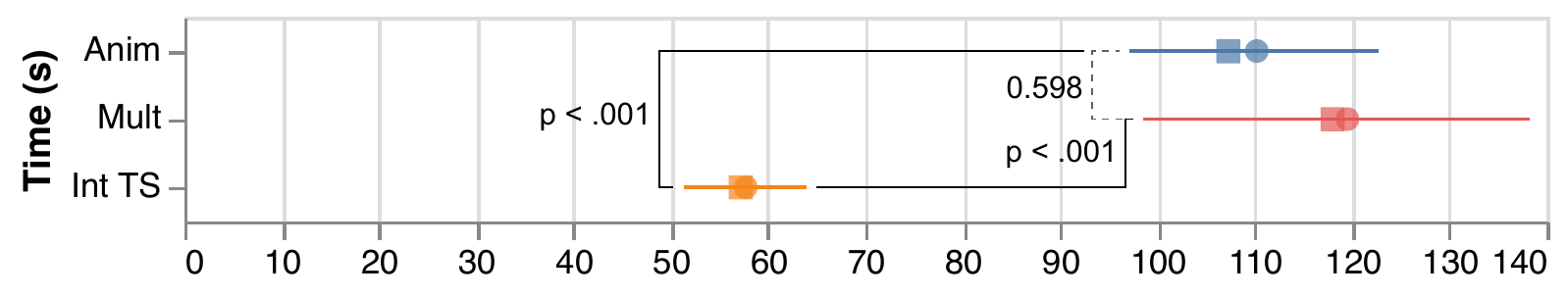}
    \caption{Experiment 1 completion time (s) in seconds for the near (top) and far (bottom) conditions without log transform.  The dot represents the mean, square the median, and 95\% CIs shown.  Pairwise lines indicate p-values with solid lines indicating significant differences.}
    \label{fig:E1Time}
\end{figure}

\paragraph{\bf Results.} Correctness and completion time are shown in Figures~\ref{fig:E1Corr} and~\ref{fig:E1Time} respectively.  After a Holm–Bonferroni correction, we found significant differences in terms of correctness for the near condition with interactive timeslicing outperforming animation with a difference of $14$ percentage points ($W = 82$, $p < 0.001$) and small multiples with a difference of $6$ percentage points  ($W = 148$, $p=0.003$).  On the far condition, interactive timeslicing outperformed animation with a difference of $11$ percentage points ($W = 109$, $p < 0.001$) and small multiples with a difference of $5$ percentage points ($W = 143$, $p=0.002$).  No other differences were statistically significant.  In terms of completion time, we have the same pattern.  On near, interactive timeslicing outperforms animation with a difference of $42.1s$ (animation $1.61\times$ slower) ($t = 5.76$, $df = 23$, $p < 0.001$) and small multiples with a difference of $28.9s$ (small multiples $1.42\times$ slower) ($t = 4.28$, $df = 23$, $p < 0.001$). On far, interactive timeslicing outperforms animation with a difference of $52.5s$ (animation $1.91\times$ slower)  ($t = 8.04$, $df = 23$, $p < 0.001$) and small multiples with a difference of $61.8s$ (small multiples $2.07\times$ slower) ($t = 8.01$, $df = 23$, $p < 0.001$).


\begin{figure}
    \centering
    \includegraphics[width=0.5\textwidth]{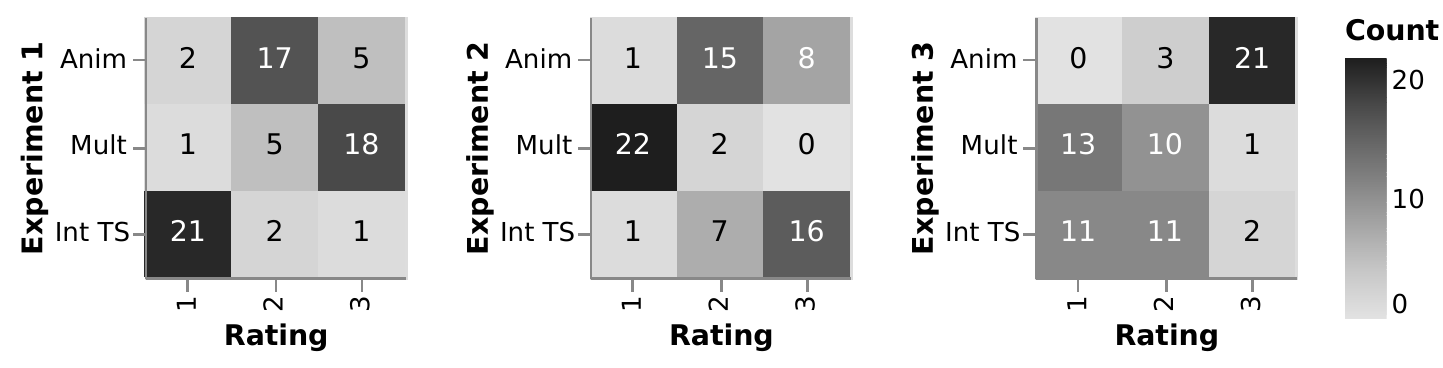}
    \caption{Participant interface rankings.  Cells annotated by number of votes and darker cells correspond to more votes.  Rating is 1st, 2nd, or 3rd.}
    \label{fig:qualresponses}
\end{figure}

\paragraph{\bf Participant Survey.}
After completing the experiment, participants ranked each condition on a scale of 1 (best) to 3 (worst) to indicate their preference for conditions to complete the given task type.
These ranking responses are in Fig.~\ref{fig:qualresponses}. For this experiment $87.5\%$ of participants indicated that they preferred the interactive timeslicing, with interactive animation second, and small multiples as the least preferred option. The primary criticism of small multiples was that, for tasks involving the far in time condition, it was impossible to put both timeslices on screen simultaneously in order to directly compare them. Participants found it challenging to remember edge positions and edge presence when they had to scroll through many intermediate representations to reach the comparison target.

The memory cost was also probably a factor in the ranking of animation. However, this was less pronounced as the interactive animation interface made it simpler to switch backwards and forwards in time without having to view intervening timeslices.

\paragraph{\bf Summary and Discussion.}
For both the near and far conditions, interactive timeslicing outperforms animation and small multiples in terms of correctness and completion time, confirming our conjectures for all research questions.  There were no other significant differences found in the experiment.  Interactive timeslicing was primarily designed for dynamic graphs spanning a long interval of time and specifically for the case of comparing distant points in time, as required by the task.  Animation incurs higher interaction costs by requiring interaction to play the animation back and forth between the distant time periods. Small multiples requires the participant to scroll back and forth between the distant time periods.  In addition, both animation and small multiples require participants to remember the structure at distant time points, whereas interactive timeslicing is able to show representations simultaneously and side-by-side on screen.  It seems that the added interaction cost of interaction timeslicing is offset by this benefit.  It is important to note that we did not see a significant difference between animation and small multiples for either near or far.  This could be due to the fact that neither of these interfaces were designed with long time series in mind.  Thus, on the task of comparing distant points in time, these interfaces were too taxing on memory and interaction, which dominated the result.

%% file: sections/06-Experiment2.tex
\section{Exp. 2: Graph Structure Changes Over a Time Interval}
This experiment tests how interactive timeslicing performs over a time interval when examining a change in graph structure.  A time interval is several consecutive timeslices.  In terms of our research questions, we conjectured that interactive timeslicing would not perform as well on this task type because it is mainly designed for time points and not continuous intervals ({\bf Q1} and {\bf Q2}). We also felt that the performance of interactive timeslicing would decrease with far trials because a longer interval of time needed to be considered and more information needed to be remembered ({\bf Q3}).  

\paragraph{\bf Task and Procedure.}
The on-screen prompt provided to participants for experiment two was: `select a timeslice between the red lines where the cluster of purple nodes is most dense.'

\begin{figure}
    \centering
        \includegraphics[width=0.48\linewidth]{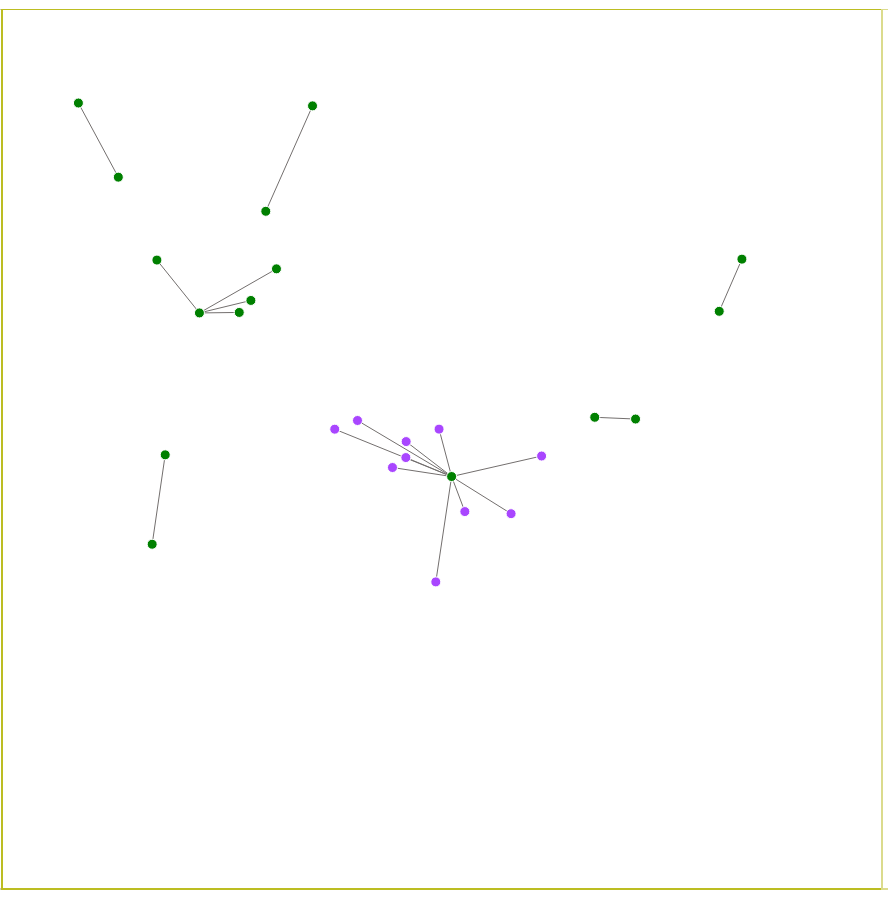}
    \caption{An example of a vignette with a cluster of purple nodes, as shown to participants during Exp. 2}
    \label{fig:task2-purplenode-example}
\end{figure}

The video figure in the supplementary material illustrates how participants answered this task on all three interfaces.  Participants saw a pair of red lines on interface timelines (see Figure~\ref{fig:interface-all}, (A), on all interfaces), indicating the beginning and end of the time interval of interest. Successful completion of the task involved investigating every six hour timeslice within this interval to identify the timeslice in which there were the highest number of connected purple nodes (see Figure~\ref{fig:task2-purplenode-example} for an example).

When participants were confident that they had found the correct answer, they would tap a button above the timeline to switch into `answer entry' mode. In this mode, participants would slide a green time window (the fixed size of one timeslice) to the position on the timeline containing their answer.

\paragraph {\bf Measurements.} Time (s) and accuracy were measured for this experiment.  Accuracy involved comparing the number of edges between purple nodes in the selected timeslice ($n_s$) to the timeslice where the number of edges is a maximum (the correct answer) ($n_a$).  In order to do this, we use the following measure:
\begin{equation}
 c = 1 - \frac{n_a - n_s}{n_a}
 \label{ceq}
\end{equation}

The value of this measure is $1$ when the correct answer is selected and diminishes to zero with a window of fewer and fewer edges between the purple nodes.

\begin{figure}[t]
    \centering
        \includegraphics[width=0.48\textwidth]{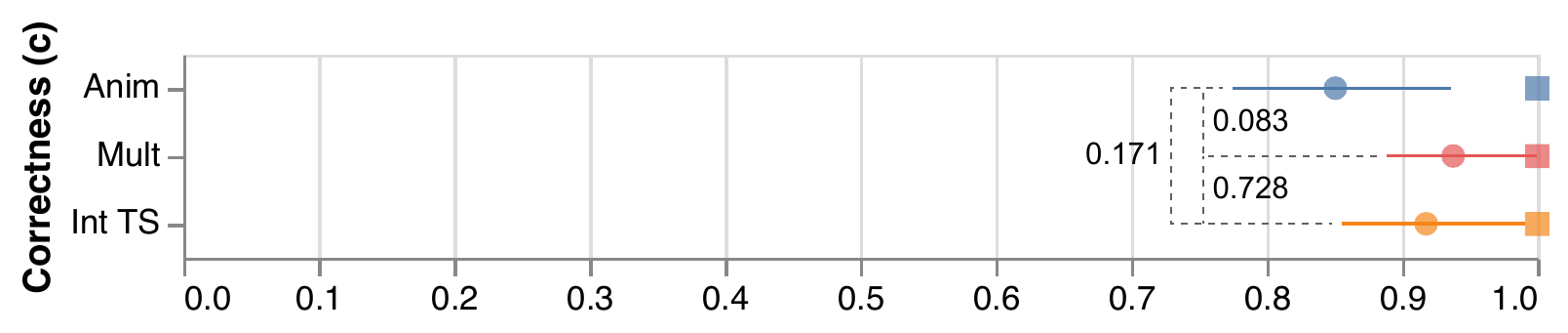}
        \includegraphics[width=0.48\textwidth]{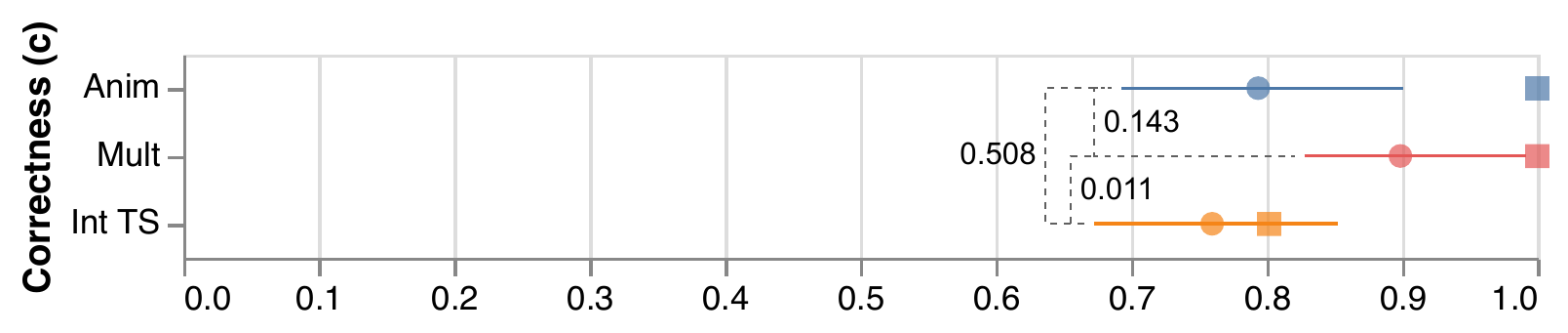}
    \caption{Experiment 2 correctness (Equation~\eqref{ceq}) for the near (top) and far (bottom) conditions.  The dot represents the mean, square the median, and 95\% CIs shown.  Pairwise lines indicate p-values with solid lines indicating significant differences.}
    \label{fig:E2Corr}
\end{figure}

\begin{figure}
    \centering
        \includegraphics[width=0.48\textwidth]{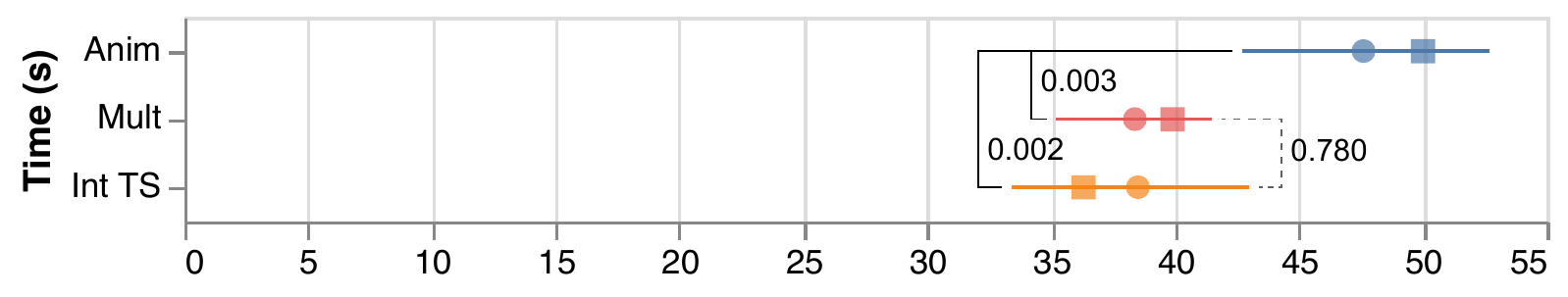}
        \includegraphics[width=0.48\textwidth]{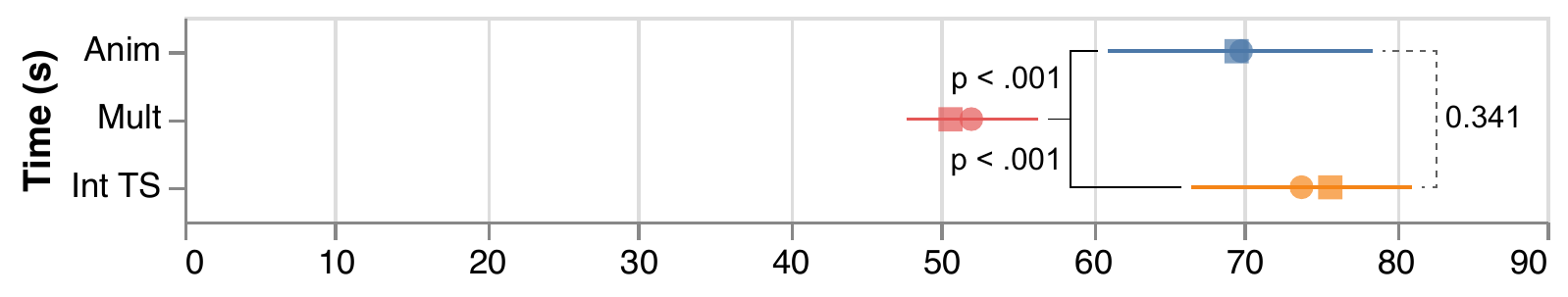}
    \caption{Experiment 2 completion time in seconds for the near (top) and far (bottom) conditions without log transform.  The dot represents the mean, square the median, and 95\% CIs shown.  Pairwise lines indicate p-values with solid lines indicating significant differences.}
    \label{fig:E2Time}
\end{figure}

\paragraph {\bf Results.} Correctness and completion time are shown in Figures~\ref{fig:E2Corr} and \ref{fig:E2Time} respectively.   After a Holm–Bonferroni correction, we found no significant differences in correctness between interfaces in neither near nor far conditions.  In terms of completion time on near, small multiples outperformed animation by $9.3s$ (animation $1.24\times$ slower) ($t = 3.31$, $df = 23$, $p=0.003$) and interactive timeslicing outperformed animation by $9.1s$ (animation $1.24\times$ slower) ($t = 3.54$, $df = 23$, $p=0.002$).  On far, small multiples outperformed both animation $17.8s$ (animation $1.34\times$ slower)  ($t = 4.01$, $df = 23$, $p < 0.001$) and interactive timeslicing $21.8s$ (interactive timeslicing $1.42\times$ slower) ($t = -6.67$, $df = 23$, $p < 0.001$).  No other significant differences were found.


\paragraph{\bf Participant Survey.}
After completing the experiment, participants ranked each condition on a scale of 1 (best) to 3 (worst) to indicate their preference for conditions to complete the given task type.
These ranking responses can be seen in Fig.~\ref{fig:qualresponses}. For this experiment, $92\%$ of participants preferred small multiples for completing tasks of this type. Animation ranked second and $67\%$ of participants ranked interactive timeslicing as the worst. 

There was widespread frustration among participants while using the interactive timeslicing condition during this experiment. The comparison of a number of time points within a time interval required much interaction effort. A large amount of very precise interactions were required to position and re-position the timeslices; in contrast, small multiples simply required scrolling the vignette display area while looking at the screen. For this task type interactive timeslicing was also vulnerable to `fat finger'~\cite{siek_fat_2005} problems, where participants aimed to carry out one operation but accidentally triggered a different one due to imprecise touching of the screen. A common example occurred when a participant tried to move a time window selector but instead activated a resize operation by selecting a handle for the selected time window. The participant then had to return the time window to its previous size and attempt to carry out the re-positioning operation again.  

\paragraph{\bf Summary and Discussion.}
For this experiment, we found no significant differences in correctness.  Therefore, we have no evidence that any interface was more accurate than another,
but some of the interfaces were more efficient.  When the time interval is smaller, we can conclude that animation is slower than both small multiples and interactive timeslicing.  Interactive animation is the only one of these interfaces where all timeslices of the time interval must be remembered in order to compare them.  For both small multiples and interactive timeslicing some of the representation of the time interval can be offloaded to the interface as multiple timeslices are shown.  When the time interval is wider, we can conclude that small multiples is faster than both animation and interactive timeslicing.  As the interval of time considered increases, interactive timeslicing is more strongly affected as this technique is based on individual timeslices.  The implementation of small multiples from this experiment provides a more natural interaction with a time interval as all timeslices must be contiguous.  

%% file: sections/07-Experiment3.tex
\section{Exp. 3: Attribute Changes at Points in Time}
\label{sec:experiment3}
This experiment tests performance when reading attribute values at multiple, disjoint, time points.  In terms of research questions, we thought that interactive timeslicing would perform well here ({\bf Q1} and {\bf Q2}).  We felt that this difference would increase for the far level ({\bf Q3}).

\begin{figure}
    \centering
    \includegraphics[width=0.48\textwidth]{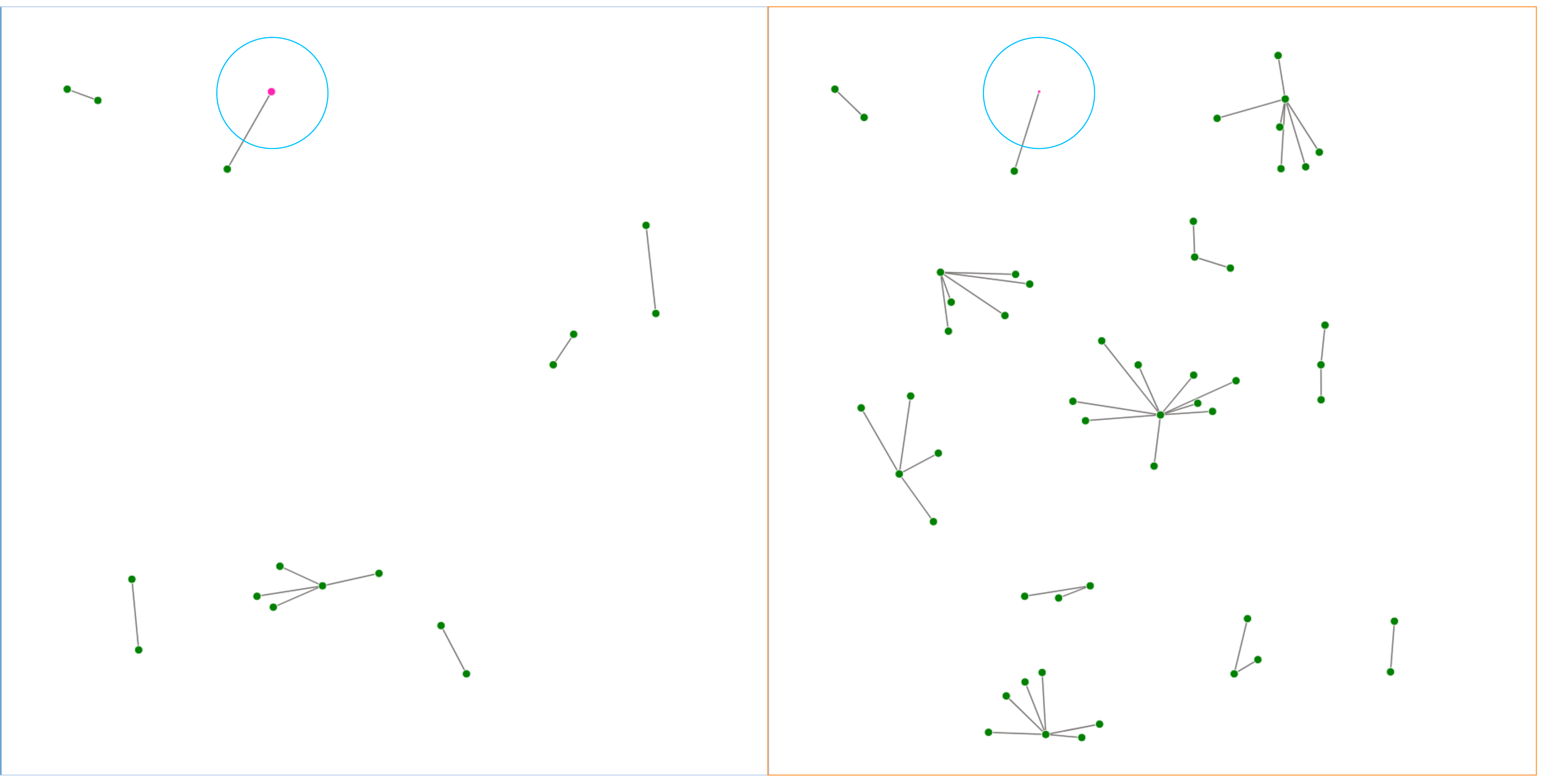}
    \caption{Two vignettes, as seen during task 3, with a standard size pink node (left) and a small size pink node (right). Pink nodes are highlighted with a blue circle for the purposes of readability, but this was not present on the experiment interfaces.}
    \label{fig:task3-vignette-example}
\end{figure}

\paragraph{\bf Task and Procedure.}
The instruction for this third experiment was `each pair of red (start) and blue (stop) lines signify a timeslice where a pink node is present. In which timeslice is the pink node smallest?'. The video figure in the supplementary material illustrates how participants answered this task on all three interfaces. A randomly selected node, present throughout the interval of interest, was given a minimum attribute value at precisely one of those timeslices and a maximum value in all others.  Bright pink was chosen as the color of the target node for contrast reasons to minimize visual search time. The standard node size is 3.3$\times$ bigger than the size of the smaller answer node. An example of a normal pink node vs. the smallest pink node can be seen in Figure~\ref{fig:task3-vignette-example}.

The procedure for submitting answers for this task was the same as {\bf Exp 2}, with participants using a button to enter `answer mode' and moving a timeslice, of fixed size, to their answer position. 

\paragraph{\bf Measurements.}  Time (s) and accuracy were measured for this experiment.  There was only one correct answer where the attribute was at its minimum value.  Therefore, a score of $1$ was recorded for each task answered correctly and $0$ for an incorrect answer.

\begin{figure}
    \centering
    \includegraphics[width=0.48\textwidth]{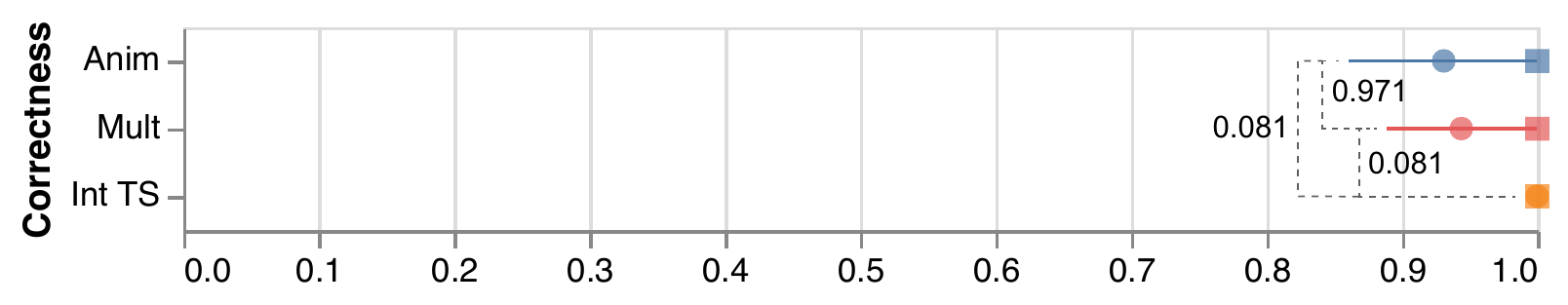}
    \includegraphics[width=0.48\textwidth]{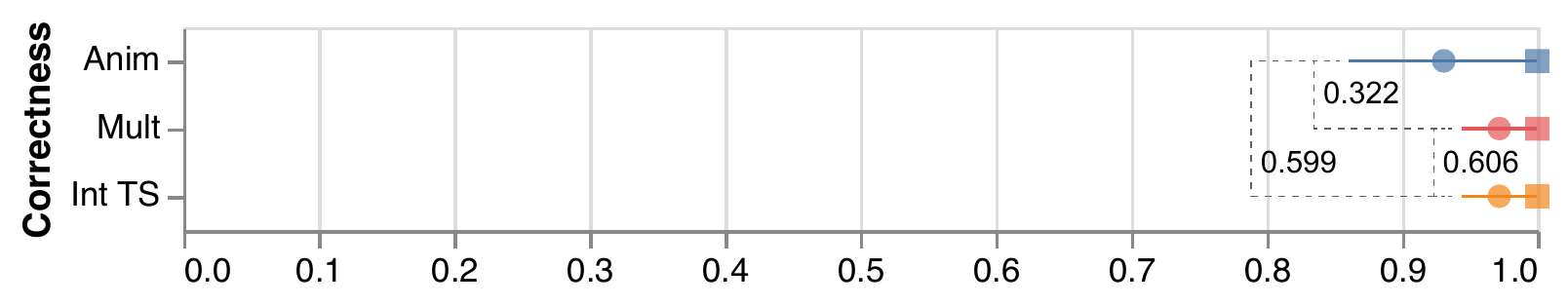}
    \caption{Experiment 3 correctness for the near (top) and far (bottom) conditions.  The dot represents the mean, square the median, and 95\% CIs shown.  Pairwise lines indicate p-values with solid lines indicating significant differences.}
    \label{fig:E3Corr}
\end{figure}

\begin{figure}
    \centering
    \includegraphics[width=0.48\textwidth]{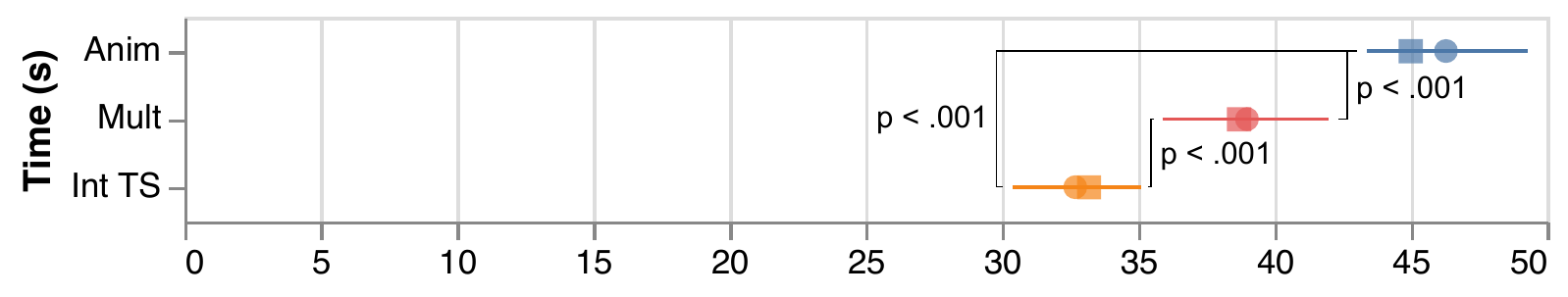}
    \includegraphics[width=0.48\textwidth]{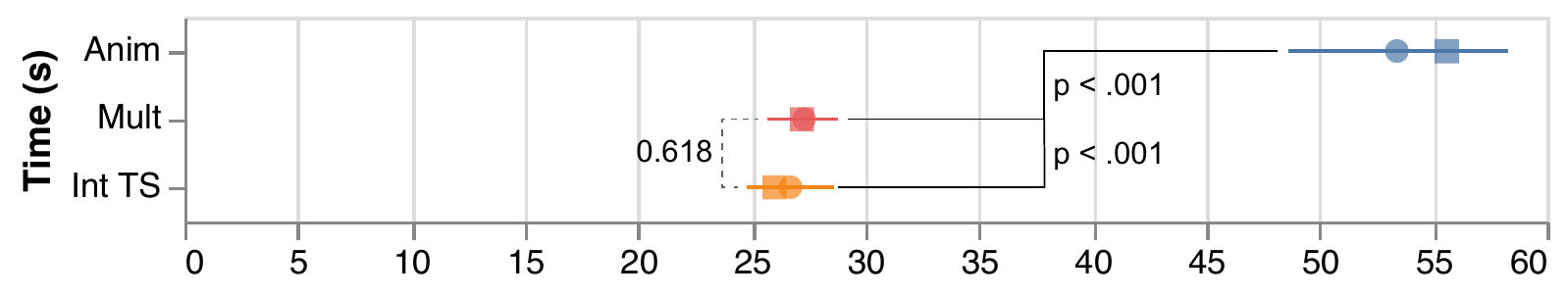}
    \caption{Experiment 3 completion time in seconds for the near (top) and far (bottom) conditions without log transform. The dot represents the mean, square the median, and 95\% CIs shown.  Pairwise lines indicate p-values with solid lines indicating significant differences.}
    \label{fig:E3Time}
\end{figure}

\paragraph {\bf Results.} Correctness and completion time are shown in Figures~\ref{fig:E3Corr} and \ref{fig:E3Time} respectively.  After a Holm–Bonferroni correction,  we found no significant differences between the interface conditions in neither the near nor the far condition.  All interfaces had median 100\% correctness for this experiment for both near and far conditions.  In terms of completion time on near, all pairwise differences were significant with interactive timeslicing outperforming both small multiples by $6.3s$ (small multiples $1.19\times$ slower) ($t = 4.45$, $df = 23$, $p < 0.001$) and animation by $13.6s$ (animation $1.42\times$ slower) ($t = 8.52$, $df = 23$, $p < 0.001$), and small multiples outperforming animation by $7.3s$ (animation $1.19\times$ slower) ($t = 4.80$, $df = 23$, $p < 0.001$).  On the far condition, small multiples outperformed animation by $26.1s$ (animation $1.96\times$ slower) ($t = 12.7$, $df = 23$, $p < 0.001$) and interactive timeslicing outperformed animation by $26.7s$ (animation $2.00\times$ slower) ($t = 11.8$, $df = 23$, $p < 0.001$).  The remaining pairwise difference between interactive timeslicing and small multiples was not significant.


\paragraph{\bf Participant Survey.}
After completing the experiment, participants ranked the interfaces on a scale of 1 (best) to 3 (worst) in order of preference to complete the given task type. For full results see Fig.~\ref{fig:qualresponses}.
There was no clear preference for a single interface for completing this task type. However, animation was strongly disliked with $87.5\%$ participants giving it a rank of 3. There is a relatively low interaction cost for small multiples and interactive timeslicing for this task, whereas animation required participants to scroll across the timeslices. Interactive animation also required a much higher level of concentration for this task type than the other interfaces, with participants having to identify the target node and then remember its position while animating the graph. The opposite is the case for interactive timeslicing and small multiples where all nodes and edges were always visible, meaning participants simply had to identify the pink node in each time window of interest rather than attempt to remember the previous node position.

\paragraph {\bf Summary and Discussion.}
For this experiment, no significant differences were found in terms of correctness. Hence there is no evidence supporting superior accuracy of any of the interfaces.  
In terms of completion time for near, all pairwise differences were significant with interactive timeslicing the fastest, followed by small multiples, and then animation.  As the near condition is closest to previous experiments~\cite{Archambault2016,farrugia_effective_2011}, small multiples outperforming interactive animation is consistent with this result.  For far, we did not see a difference between interactive timeslicing and small multiples where there was one for near.  However, animation is significantly slower than both interfaces.  There could be many reasons why we did not find a difference between small multiples and interactive timeslicing.  One possible interpretation is that the task is less demanding on participant memory (remembering a single node instead of a collection of edges), and thus, the difference is less pronounced.  Further experimentation is required to test this hypothesis.

%% file: sections/08-Discussion.tex
\section{General Discussion}
We first summarize the main results from the experiment based on the research questions from Section~\ref{sec:questions}. Answers to the questions vary depending on the task, with Experiment 1 and 3 showing similar patterns but different from Experiment 2.

In Experiments 1 and 3, task completion times with the different techniques (Q1) show the clear advantage of interactive timeslicing over the other techniques. We take this as indication that interactive timeslicing facilitates time navigation between discrete points, which is what the tasks from Experiments 1 and 3 have in common. There is one exception: in the far condition of Experiment 3, small multiples and interactive timeslicing showed similar completion times (i.e., we observed a difference between the near and far conditions --- Q3). We attribute this to a reduced memory requirement for that condition. Occasionally small multiples visibly outperformed animation. In accuracy (Q2), interactive timeslicing also showed a clear advantage over the other two techniques in Experiment 1, but not in Experiment 3, where accuracy was high for all interfaces and statistically indistinguishable.

Experiment 2 has very different results. Small multiples was faster than the other techniques in the far condition (Q1), although statistically indistinguishable from interactive timeslicing in the near condition. Accuracy measurements in this experiment show fairly large differences in means, but large variance in the trials prevents us from finding statistically reliable differences between interfaces in both near and far conditions (Q2). A possible explanation of why small multiples performed significantly better is that it naturally represents consecutive time windows that can be scrolled through easily.



When considering the survey data, no interface was preferred by the majority of participants for all tasks.  Instead, interface preference is task dependent. For the first and second experiments the rankings did indicate clear interface preferences for the completion of those task types; interactive timeslicing and small multiples ranked first for those experiments, respectively. In contrast to Experiment 1 and 2, there was no clear preference for a single interface for completing the third experiment.  From our own observations of Experiment 3, there seemed to be fewer `fat finger'~\cite{siek_fat_2005} interaction mistakes with interactive timeslicing than in Experiment 2. It is likely that this is because there was no requirement to move or resize a time window after the initial definition stage (barring participant error). One participant remarked that their version of a perfect interface would be small multiples with the ability to make selected vignettes appear and disappear to better facilitate side-by-side comparison of highly distant time points. 
 
Interactive animation is often slower and less preferred but with no difference in terms of correctness, meaning it is a less efficient way of finding the correct solution.  On Experiment 1, it is significantly slower than interactive timeslicing.  On Experiment 2, it is significantly slower than small multiples.  On Experiment 3, it is significantly slower than the other two interfaces.  Thus, for tasks involving time navigation as tested in these experiments, we confirm some results of other experiments on dynamic data~\cite{Archambault2016,5473226,farrugia_effective_2011,Robertson2008,Brehmer2020}.  One possible conjecture for why it is slower is that the participant has no idea where to look in the animation when undertaking an exploratory task.  Although there are some preliminary results~\cite{Robertson2008}, it remains an open question whether animation works well for explanatory tasks where a presenter can point out regions of interest for another viewer to understand.
 
Brehmer et al.~\cite{Brehmer2020} compared the efficiency of animation and small multiples on mobile phones for animated scatterplots.  The result of this previous experiment found that small multiples was usually faster than animation with no difference in correctness.  One could view our study as a somewhat analogous test on large displays with a touch interface that confirms several results comparing the interactive animation and small multiples conditions.
 

%% file: sections/09-Conclusion.tex
\section {Limitations and Future Work}\label{sec:limitations}
As with all experiments, experimental design choices cause limitations in result interpretability.  One such choice is that we prioritized counterbalancing the interfaces, and not the experiments or near/far factor, in order to reduce experimental noise.  Thus, later experiments might have had more tired participants and earlier experiments had less training.  Far tasks might have benefited from the experience of having done the near tasks first. Nevertheless, we hypothesize that it is unlikely that training or fatigue might have affected the interfaces differently.

All trials in our experiment use parts of the same dynamic network data. Although this dynamic graph is long in time with much variability in the different time points and intervals of the different trials, it is important to test other datasets in the future for the sake of generalizability and to further explore which types of structures or values might affect the different interfaces.  

In order to keep a reasonable experiment length, we only tested three task types, but understanding how these interfaces perform for a wider set of tasks (e.g., additional ones from~\cite{Andrienko_task_2005} or~\cite{kerracher_task_2015,14Ahn}) would produce further insights.  We only test our interfaces with node-link diagrams despite matrices being another popular method for representation~\cite{Beck2017}. Testing the interfaces with a wider range of graph representations and task types would ensure that results are more generalizable.  

It is also important to remark that we also made specific design decisions in the implementation of the visualization interfaces, usually to support fair comparison between techniques. For example, only four small multiples were visible at a time and scrolling was required, and the most basic form of interactive animation was used whereby individual events are controlled with a slider. Similarly, we used linear interpolation in our animations; staged animated transitions could be considered~\cite{6658746,14Chevalier}. The effects of some of these secondary design decisions have the potential to be important, but are out of scope for this work. Future work manipulating further parameters will be welcome and should compare and further extend our findings.  

It is possible that some of the advantages and disadvantages that we observed are significantly affected by the type of input (e.g., direct touch vs. indirect mouse). Although we have justified testing with touch input as the more natural way to work on large collaborative displays, indirect inputs such as computer mice or touch pads are probably still a more common way to interact with dynamic network visualizations. Hence, it is important to further investigate the role of input type in these results. We are already working on an experiment to empirically verify possible differences caused by input.

Similarly, the size of the display or the portion of the field of view that it covers could explain some of the differences that we observed. Investigating how performance with these interfaces varies will provide generalizability, but it could also offer valuable insights to design improved variants that are even better.

\section{Conclusion}
In this paper, we present the results of a series of experiments intended to formally evaluate methods to visualize dynamic networks on large touch displays. We were primarily interested in comparing interfaces for tasks that involve cross-time operations to see network structure and attribute variation.
Two of our selected interfaces, interactive animation and small multiples, are already well studied in previous literature~\cite{Robertson2008,Archambault2016,farrugia_effective_2011,5473226,Brehmer2020}. Our third selected interface, interactive timeslicing, has not previously been experimentally evaluated.  

For tasks involving comparison of specific time points, interactive timeslicing offered greater speed than interactive animation or small multiples; when the comparison was of network structure (collections of edges), there was also an important difference in accuracy. In several instances, small multiples was also better than interactive animation.

For navigating time intervals, small multiples is faster than both interactive animation and interactive timeslicing when time intervals are larger (the far condition). Small multiples has the advantage that it naturally represents contiguous time intervals, whereas interactive timeslicing requires the participant to create and move each of these timeslices individually. Interactive timeslicing was also ranked by participants as the worst interface for completing tasks of this type. Whilst completing an exploration of time intervals some participants struggled with `fat finger'~\cite{siek_fat_2005} problems; time windows would be accidentally re-sized rather than moved. Further refinement of interactive timeslicing would help resolve the issues of distinguishing between these interactions. 

Due to the lack of existing evaluations for interactive timeslicing, assessing interface performance with solo participants is a necessary starting point. However, large displays are commonly used in collaborative settings and previous evaluations have shown the value of collaboration for graph exploration~\cite{7516722}. With this in mind, future experiments evaluating the usability of these interfaces for collaborative situations is vital. 

